\documentclass{cpbtex}
\usepackage{graphicx}

\begin{document}
%*\begin{CJK*}{GBK}{song}

\title{Magnetoresistance hysteresis in the superconduting state of Kagome CsV$_3$Sb$_5$}

% Title should be concise; avoid abbreviations if possible; and not begin with `A', `An', `The', or `Study on'.

\author{Tian Le$^{1,2,}$\thanks{Corresponding author. E-mail:letian@westlake.edu.cn}, \ Jinjin Liu$^{3,4}$, \ Zhiwei Wang$^{3,4,5}$, \ and \ Xiao Lin$^ {1,2,}$\thanks{Corresponding author. E-mail:linxiao@westlake.edu.cn}\\
$^{1}${Institute of Natural Sciences, Westlake Institute for Advanced Study,}\\
 {Hangzhou 310024, P. R. China}\\  % The line break was forced via \\
$^{2}${Key Laboratory for Quantum Materials of Zhejiang Province, Department of Physics,}\\
{School of Science and Research Center for Industries of the Future,}\\
{Westlake University, Hangzhou 310030, P. R. China}\\ % The line break was forced via \\
$^{3}${Centre for Quantum Physics, Key Laboratory of Advanced Optoelectronic Quantum}\\
{Architecture and Measurement (MOE),}\\
{School of Physics, Beijing Institute of Technology, Beijing 100081, China}\\
$^{4}${Beijing Key Lab of Nanophotonics and Ultrafine Optoelectronic Systems,}\\
{Beijing Institute of Technology, Beijing 100081, China}\\
$^{5}${Material Science Center, Yangtze Delta Region Academy of Beijing}\\
{ Institute of Technology, Jiaxing 314011, China}
} % The line break was forced via \\

% 1. For Chinese authors, the name in Chinese characters should also be given. For example, Gang Liu(Áõ¸Õ), Xiao-Ming Li(ÀîÏþÃ÷)
% 2. Please ensure that every author approves the submission of the manuscript
% 3. Abbreviations should not be used in the affiliations

\date{\today}
\maketitle
\begin{abstract}
\noindent
The  hysteresis of magnetoresistance observed in superconductors is of great interest due to its potential connection with unconventional superconductivity. In this study, we perform electrical transport measurements on Kagome superconductor CsV$_3$Sb$_5$ nanoflakes and uncover unusual hysteretic behaviour of magnetoresistance in the superconducting state. This hysteresis can be induced by applying either a large DC or AC current at temperatures ($T$) well below the superconducting transition temperature ($T_{\rm c}$). As $T$ approaches  $T_{\rm c}$, similar weak hysteresis is also detected by applying a small current. Various scenarios are discussed, with particular focus on the effects of vortex pinning and the presence of time-reversal-symmtery-breaking superconducting domains. Our findings support the latter, hinting at chiral superconductivity in Kagome superconductors.

\end{abstract}


\textbf{PACS:} 74.78.-w;74.25.F-;85.25.-j;74.25.-q;%no more than four PACS codes should be provided: check https://cpb.iphy.ac.cn/UserFiles/File/PACS2010Regular-Edition.pdf

\section{Introduction}
\noindent

Hysteresis is a phenomenon whereby the response of a system to an external disturbance, such as magnetic field ($H$), electric field, temperature ($T$) or mechanical stress, depends on the history of applied perturbation. Understanding of hysteresis is beneficial for designing materials of improved quality and predicting novel properties \cite{PhysRevB.105.014405, PhysRevB.109.104414, BOS, STO, stress, PhysRevLett.128.036401}. In superconductors, the presence of hysteretic magnetoresistance has attracted considerable attention, because of its relation to several intriguing dynamics, including magnetic vortices, ferromagnetism, chiral superconducting domains and spin-triplet paring symmetry \cite{ PhysRevB.47.470, PhysRevB.67.134506, FeTe, semenov2019dissipation, PhysRevLett.107.056802, KTO-ETO, CoSi2, Bi/Ni, Sr2RuO4, RbV3Sb5}. Relevant investigation will provide valuable insights for the development of novel superconducting materials and various quantum technologies, ranging from energy transmission to quantum computation.

As a member of Kagome superconductors, CsV$_3$Sb$_5$ (CVS) has garnered significant interest due to its unique electronic properties and diverse quantum orders of matter~\cite{CVSreview1, CVSreview2}. In addition to the fully gapped superconductivity~\cite{mu2021s, duan2021nodeless, roppongi2023bulk, zhong2023nodeless}, a pair density wave was detected by scanning tunneling microscopy, showing remarkable resemblance to that in cuprates~\cite{CVSPDW}. Moreover, multicharge flux quantization was observed in mesoscopic CVS ring devices,  suggesting the possibility of higher charge superconductivity \cite{PhysRevX.14.021025}. Numerous theoretical studies have predicted chiral superconductivity in CVS, given the Kagome lattice formed by vanadium atoms and breaking of time-reversal symmetry (TRS) in the normal state~\cite{ko2009doped, zhao2021cascade, mielke2022time, guo2022switchable,farhang2023unconventional, gupta2022two, yu2021evidence, guguchia2023unconventional}. Recently, a technique combining the zero-field superconducting diode effect and interference patterns was employed to demonstrate the presence of TRS-breaking superconducting order with dynamic domains in CVS~\cite{le2024superconducting}.

In this study, we performed electrical transport measurements on CVS nanoflakes under various DC bias current ($I_\mathrm{DC}$), AC excitation current ($I_\mathrm{AC}$) and $T$. In the superconducting state, an intriguing hysteresis of magnetoresistance within a field range of $\pm$ 20 mT was observed when sweeping the field in opposite directions. In this context, the potential roles of vortex pinning effects and chiral superconducting domains were discussed in elucidating the origin of the observed phenomena.

\section{Experimental methods}
\noindent

Single crystals of CsV$_3$Sb$_5$ were grown through flux methods by using Cs (purity 99.8\%) bulk, V (purity 99.999\%) pieces and Sb (purity 99.9999\%) shot as the precursors and Cs$_{0.4}$Sb$_{0.6}$ as the flux agent. The starting elements were placed in an alumina crucible and sealed in a quartz ampoule in an argon-filled glove box.  The ampoule was then gradually heated up to 1000~\textcelsius ~ in 200 h and held at that temperature for 24 h in an oven. It was subsequently cooled down to 200 \textcelsius~at a rate of 3.5 \textcelsius/h. The resulting product was immersed in deionized water to remove the flux. Finally, shiny CsV$_3$Sb$_5$ crystals with hexagonal shape were obtained. CsV$_3$Sb$_5$ nanoflakes were mechanically exfoliated from the bulk crystals with a thickness exceeding 40 nm, using Nitto blue tape and transferred onto silicon substrates capped with 300 nm SiO$_2$. The contacts were deposited with Ti (5 nm)/Au (80 nm) via electron beam evaporation. In order to improve the quality of the contacts,  the flakes were cleaned by Ar plasma prior to deposition.

The transport measurements were performed in Quantum Design physical property measurement system (PPMS). The resistance ($R$) and differential resistance (d$V$/d$I$) for three devices (D1-D3) were measured by a standard four-terminal method. d$V$/d$I$ in the positive ($I_\textrm{+}$) and negative ($I_\textrm{-}$) bias current regimes was collected by sweeping ($I_\mathrm{DC}$) from 0 to $I_\textrm{max+}$ and $I_\textrm{max-}$, respectively. $I_\mathrm{DC}$ was supplied by a current source meter (Keithley 2450). A lock-in amplifier (Stanford Research, SR830) combined with a 100 k$\Omega$ buffer resistor was used to offer small $I_\mathrm{AC}$ ($11~\mathrm{Hz}-173$~Hz, $1~\mu\mathrm{A}-50~\mu$A) to detect the differential resistance (d$V$/d$I$ = $V_\textrm{AC}$/$I_\textrm{AC}$).

\section{Results and discussion}
\noindent

\begin{center}
\includegraphics[width=0.8\textwidth]{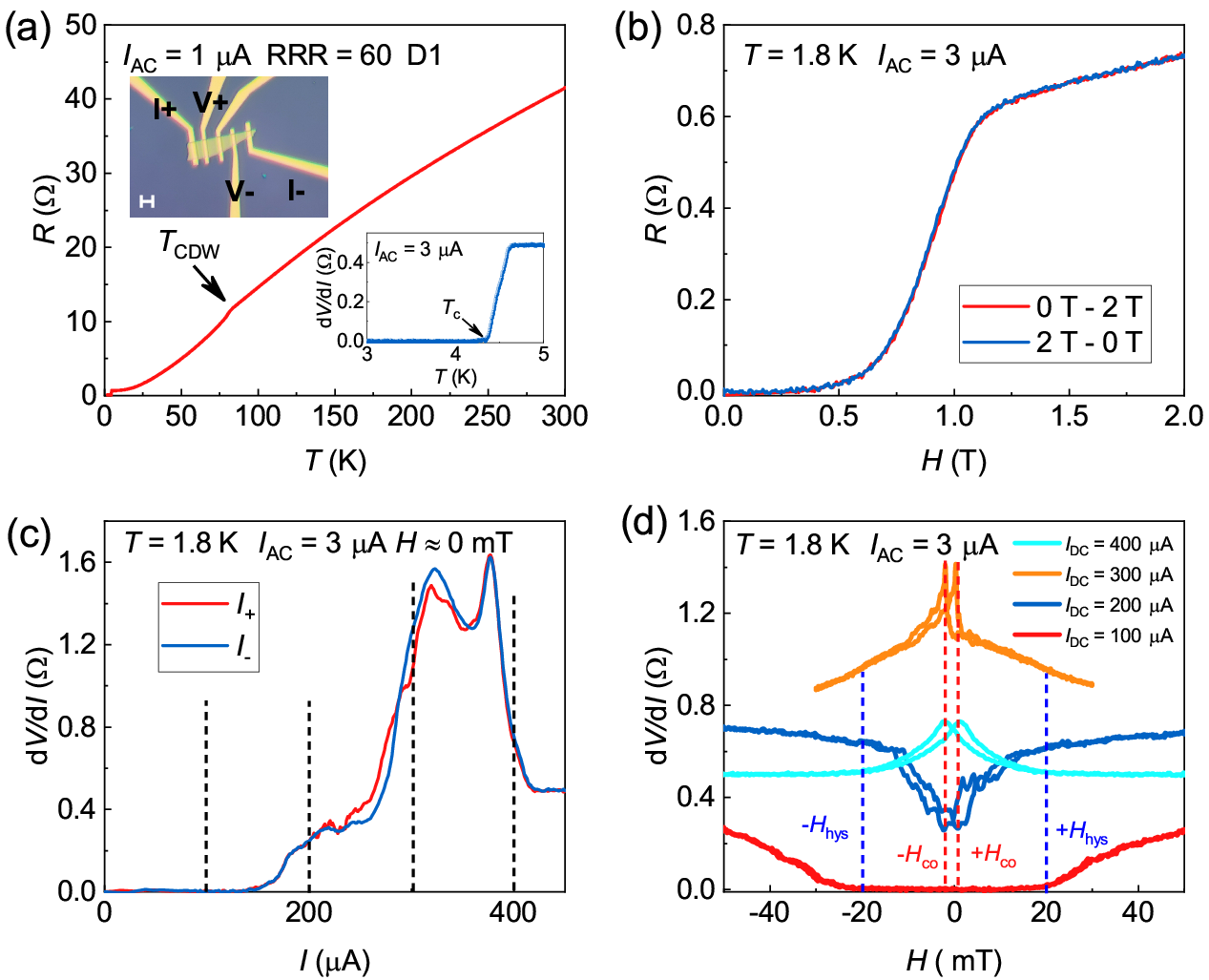}\\[5pt]  % insert figure
\parbox[c]{15.0cm}{\footnotesize{\bf Fig.~1.} (a) Temperature dependent resistance for D1 from 300 K to 1.8 K. The lower inset is the enlarged curve around $T_\mathrm{c}$. The upper inset shows the optical image of D1 with a scalar bar of 4 $\mu$m.
(b) Magnetic field dependent resistance for D1 at 1.8 K. The red and blue curves represent the field swept from 0 to 2 T and 2 to 0 T, respectively.
(c) The differential resistance as a function of current bias at 1.8 K for D1. The red(blue) curve represents the current swept from 0 $\mu$A to positive (negative) bias. The dashed lines mark the bias current at 100 $\mu$A, 200 $\mu$A, 300 $\mu$A and 400 $\mu$A, respectively. (d) Field dependent differential resistance at four bias current marked in (c). All curves are swept with a field cycling beginning at negative field. The blue dashed lines denote the onset of hysteresis. The red dashed lines denote the coercive field. The field sweep rate is 0.3 mT/s. }
\end{center}

Figure. 1(a) illustrates $R$ versus $T$ for D1 in a full temperature range, which indicates high sample quality with a residual-resistance-ratio (RRR) of 60. The transitions of charge density wave (CDW) and superconductivity occur at $T_{\rm CDW}$ = 82 K and $T_{\rm c}$ = 4.3 K, respectively. Fig. 1(b) shows the magnetoresistance measured at $T=1.8$~K and  $I_{\rm AC}$ = 3 $\mu$A  by sweeping $H$ in a large field range, i.e. from 0 T to 2 T (red) and back from 2 T to 0 T (blue) respectively. It clear that two curves collapse without detectable hysteresis. As $H$ increases, the fast enhancement of finite resistance could be attributed to vortex depinning effect. The absence of hysteresis observed in Fig. 1(b) suggests that this pinning effect is not prominent \cite{samukawa2024observation, PhysRevB.47.470, PhysRevB.67.134506, FeTe, semenov2019dissipation}.

In Fig. 1(c), we present d$V$/d$I$ as a function of $I_{\rm DC}$ at 1.8 K, where the superconducting state sustains up to 420 $\mu$A. The discrepancy between positive ($I_{\rm +}$) and negative ($I_{\rm -}$) sweeping curves arises from the superconducting diode effect, as reported in our previous work\cite{le2024superconducting}. The presence of multiple transition peaks implies signatures of superconducting domains, likely chiral domains~\cite{le2024superconducting}.
Figure~1(d) present d$V$/d$I$ versus $H$ for $I_\mathrm{DC}=100-400~\mu$A, corresponding to the vertical lines in Fig. 1(c). At $I_{\rm DC}=200-400~\mu$A, these curves exhibit a notably bow-tie-shaped hysteresis loop within the superconducting state. This resembles the behavior in ferromagnetic superconductors~\cite{PhysRevLett.107.056802, KTO-ETO}, where the positions of dips (peaks) in the d$V$/d$I$ curves define the coercive field $H_\mathrm{co}$. The hysteresis merges at slightly higher field ($\pm20$~mT), corresponding to the full polarization of domains, denoted as $H_{\rm hys}$.  However, it is well-established that there is no ferromagnetism in CVS~\cite{CVSreview1, CVSreview2}. On the other hand, in the superconducting mixed state, vortex depinning occurs once its Lorentz force exceeds the pinning force caused by defects, with the former proportional to the product of the external field and current density ($J$)~\cite{tinkham2004introduction}. Therefore, the higher the current, the lower the depinning field. This is in stark contrast with the observation in Fig.~1(d), wherein even as $I_\mathrm{DC}$ increases, both $H_\mathrm{co}$ and $H_{\rm hys}$ remain roughly unchanged.

In a chiral superconductor, it is energetically favorable to form domain structures between degenerate superconducting phases of opposite chiralities. Similar to ferromagnetic domains, these superconducting domains are pinned by disorder or defects. The field strength, exceeding the pinning force, can switch the chirality of domains, resulting in hysteresis during field cycling~\cite{bouhon2010influence}. This field is independent of the applied current, consistent with the observation in Fig. 1(d). At $I_{\rm DC}$ = 100 $\mu$A, the hysteresis is negligible since the resistance is zero within $\pm H_{\rm hys}$. Subsequent data will extend the discussion in the frame of chiral domains.

\begin{center}
\includegraphics[width=0.5\textwidth]{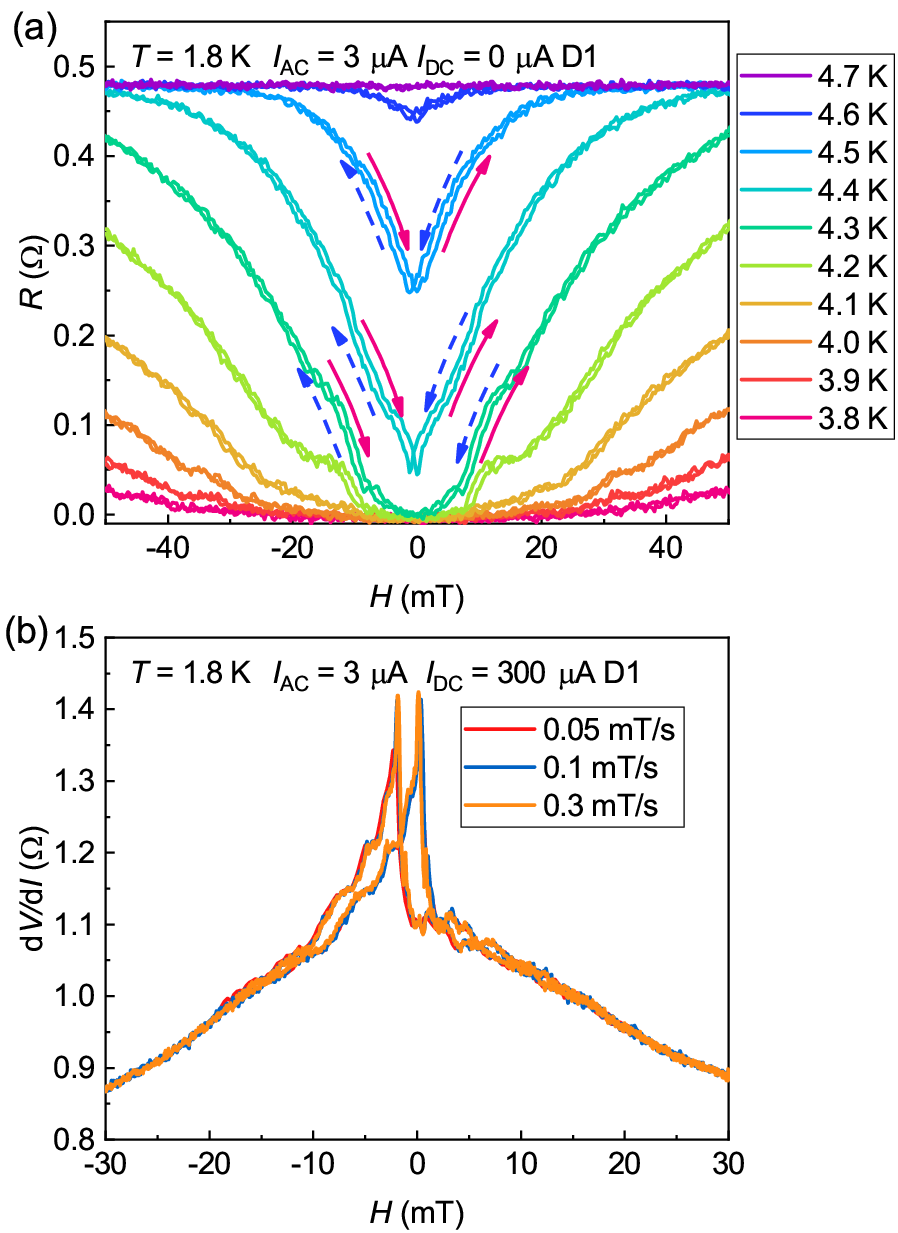}\\[5pt]
\parbox[c]{15.0cm}{\footnotesize{\bf Fig.~2.}   % figure caption
(a) Field dependent resistance for D1 at various $T$ around $T_\mathrm{c}$. The field sweep rate is 0.3 mT/s. (b) Field dependent resistance for D1 at $T=1.8$~K, $I_{\rm DC} =300~\mu$A and $I_{\rm AC}=3~\mu$A  with the field sweep rates of 0.05 mT/s, 0.1 mT/s and 0.3 mT/s, respectively.
}
\end{center}

In Fig. 2(a), we examine the field-dependent resistance at $I_\mathrm{DC}=0$ for device D1 by varying $T$. At low $T$, the resistance is nearly zero, displaying no detectable hysteresis when the field cycles between $\pm$50 mT. As $T$ approaches $T_\mathrm{c}$ and above, $R$ becomes finite and an observable bow-tie-shaped hysteresis loop emerges, similar to that in Fig. 1(d). However, the width of the hysteresis, for instance indicated by $2H_\mathrm{co}$ $\sim$ 1.4 mT, becomes much smaller than the value of $\sim$ 3.2 mT in Fig. 1(d). This weaker hysteresis can be ascribed to the reduction of the domain pinning force due to strengthened thermal fluctuation as $T$ rises, which aligns with the chiral superconducting domain scenario.  At even higher $T$ (in the normal state), this hysteresis disappears implying that it is inherent to the superconducting state. It is worth noting that vortices have been depinned at $T>T_{\rm c}$, making them irrelevant here.

\begin{center}
\includegraphics[width=0.5\textwidth]{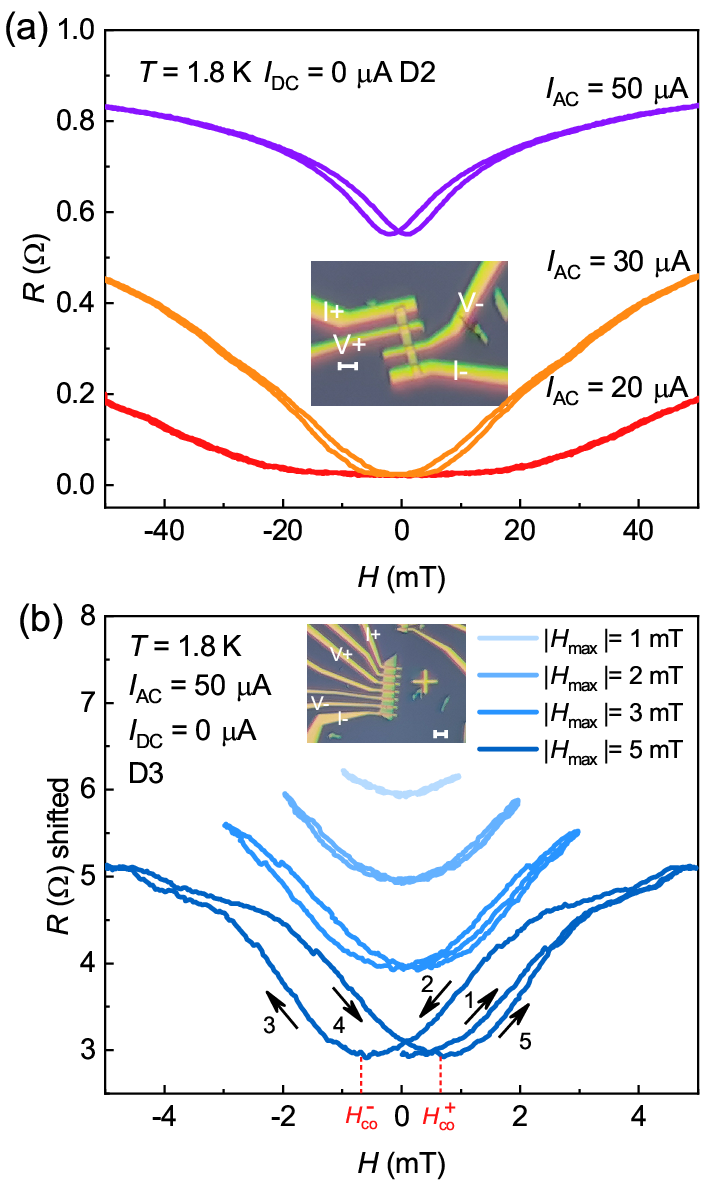}\\[5pt]
\parbox[c]{15.0cm}{\footnotesize{\bf Fig.~3.}   % figure caption
(a) Field dependent resistance for D2 at various AC excitation $I_{\rm AC}$. The inset shows an optical image of D2 with a scalar bar of 2 $\mu$m.(b) Field dependent resistance for D3 with different field sweep ranges at $I_{\rm AC}$ = 50 $\mu$A and $I_{\rm DC}$ = 0 $\mu$A. The numbers (1-5) represent the sweep protocols in sequence. The red dashed lines denote the coercive field. The inset is an optical image of D3 with the scalar bar of 6 $\mu$m. }
\end{center}

We note that artificial hysteresis may occur such that the field sweep rate exceeds the data collection rate of the computer in the measurements of $R$-$H$ curves. To check this, we measured $\textrm{d}V/\textrm{d}I$ versus $H$ for D1 by varying the field sweep rates from  0.05 mT/s to 0.3 mT/s, as shown in Fig.~2(b). All curves exhibit hysteresis and collapse with each other. It seems insensitive to the sweep rate, ruling out this possibility. Moreover, hysteresis can also be induced by heating effects from the field sweep around 0 mT. In this case, the characteristics of hysteresis would depend on the heating power related to the sweep rate. This is clearly not the case here. Therefore, the hysteresis observed in our devices is unlikely to be artificial.

Similar hysteresis can be reproduced in other CVS devices. In Fig. 3(a), we inspect the $R$-$H$ curves with different $I_\mathrm{AC}$ at 1.8 K and $I_{\rm DC}=0~\mu$A for D2. As $I_{\rm AC}$ increases, the depinning and subsequent motion of vortices generate finite resistance in a magnetic field. At $I_{\rm AC}$ = 20 $\mu$A, the resistance remains small with marginal hysteresis. As $I_{\rm AC}$ increases from 30 to 50 $\mu$A,  the hysteresis of magnetoresistance becomes increasingly remarkable, which resembles the data collected by varying $I_\mathrm{DC}$ or $T$ in Fig.~1 and 2.

In the following, let's delve into the switching characteristics of hysteresis by varying the maximum sweeping field ($H_\mathrm{max}$). The measurements were performed on D3 with $I_{\rm DC}=0~\mu$A and $I_{\rm AC}=50~\mu$A at $T=1.8$~K. During the measurements, we initially ramped $H$ from 0 to a positive field $H_\mathrm{max}^{\mathrm{+}}$, then returned to $H_\mathrm{max}^{\mathrm{-}}$ and finally back to $H_\mathrm{max}^{\mathrm{+}}$ again. As seen in  Fig. 3(b), the hysteresis in the $R$-$H$ curve is indetectable at $\left|{H_{\rm max}}\right|$ = 1 mT. As $\left|{H_{\rm max}}\right|$ increases to 2 mT, a discernible hysteresis emerges. Further increasing $\left|{H_{\rm max}}\right|$ results in a pronounced hysteresis, as observed for $\left|{H_{\rm max}}\right|=3$ and $5$~mT. This hysteresis showcases a retarded manner, i.e. the $R$-$H$ curves shifts to negative (positive) as the field sweep from the $H_\mathrm{max}^{\mathrm{+}}$ ($H_\mathrm{max}^{\mathrm{-}}$).
This phenomenon is a typical feature of domain switching with $H_\mathrm{co}$ marked by the vertical dashed lines in Fig.~3(b). At $H_\mathrm{max}\approx H_\mathrm{co}$, the chirality of  domains is minimally switched during the field cycling, resulting in the absence of hysteresis. As $H_\mathrm{max}$ rises, an increasing number of domains are switched in both sweeping protocols towards $H_\mathrm{max}^{\mathrm{\pm}}$ and leads to a more conspicuous bow-tie-shaped hysteresis, as seen in the curves for $\left|{H_{\rm max}}\right|=5$~mT.

Note that in several candidates of chiral superconductors, such as Sr$_2$RuO$_4$ \cite{Sr2RuO4} and Bi/Ni bilayer~\cite{Bi/Ni}, an "advanced" form of hysteresis has been observed in the transport properties (e.g. superconducting critical current) of their Josephson junction or SQUID devices. This feature is thought to arise from the response of chiral domains to the interaction between the Meissner screening supercurrent and the change in the applied field. This phenomenon requires the system to be in the Meissner state \cite{bouhon2010influence, Bi/Ni}. In contrast, in our case, detecting the hysteresis is challenging in a Meissner state, since it will be hidden by zero resistance. A large DC or AC current is required to stimulate the vortex motion and generate finite resistance in the mixed state.

The aforementioned hysteresis in CVS is properly interpreted by the model involving the pinning of TRS-breaking domains in the superconducting state. This aligns with recent literature demonstrating a TRS-breaking order parameter with dynamic superconducting domains in CVS \cite{le2024superconducting}.  Previous theoretical studies have also proposed a chiral superconducting phase based on the Kagome lattice of CVS \cite{PhysRevB.106.174514, wu2021nature, yu2012chiral}. Therefore, the observed hysteretic magnetoresistance may serve as an indirect evidence for the possible existence of chiral superconductivity.

\section{Conclusion}

In summary, we observe a retarded hysteresis of magnetoresistance in the superconducting mixed state of CVS. Before reaching a final conclusion, let's extend the discussion of the possible explanation involving strong vortex pinning effect. First of all, strong pinning generally exists in inhomogeneous superconductors and may lead to an ``advanced'' feature of hysteresis in the mixed state, as accounted for by a two-level critical-state model~\cite{samukawa2024observation, PhysRevB.47.470, PhysRevB.67.134506, FeTe, semenov2019dissipation}. This differs from our observation. And the absence of hysteresis shown in Fig. 1(b) also suggests weak vortex pinning strength in our single crystalline devices. Second, in our case, the hysteresis occurs in a finite resistance regime where vortices are already depinned by large DC or AC current. Therefore, the dynamics of chiral superconducting domains in CVS provides a more plausible explanation. Our research lays the foundation for investigating chiral superconductivity using current-tuning transport measurements.

\addcontentsline{toc}{chapter}{Acknowledgment}
\section*{Acknowledgment}
This research is supported by the China Postdoctoral Science Foundation (Grant No. 2022M722845 and No. 2023T160586). X.L. acknowledges support by the Zhejiang Provincial Natural Science Foundation of China for Distinguished Young Scholars under Grant No. LR23A040001 and the Research Center for Industries of the Future (RCIF) at Westlake University under Award No. WU2023C009. Z.W. is supported by the National Key R$\&$D Program of China (Grants Nos. 2020YFA0308800 and 2022YFA1403400) and the Beijing Natural Science Foundation (Grants No. Z210006).The authors thank the support provided by Dr. Chao Zhang from Instrumentation and Service Center for Physical Sciences at Westlake University.

\addcontentsline{toc}{chapter}{References}
\providecommand{\newblock}{}

%\end{CJK*}  %% end the Chinese environment
\end{document}